\documentclass{iopart}
\input epsf
\usepackage{amssymb}
\usepackage{amsfonts}
\usepackage{graphicx}
\usepackage{epsfig}

\def\fig{Fig. }

\begin{document}

\renewcommand{\figurename}{Figure}

\title{Electric control of collective atomic coherence in an Erbium doped solid}
\author{Ji\v{r}\'{i} Min\'{a}\v{r}$^{1}$, Bj\"{o}rn Lauritzen$^{1}$, Hugues de Riedmatten$^{1}$, Mikael Afzelius$^{1}$, Christoph Simon$^{1}$,  and Nicolas Gisin$^{1}$}
\address{$^1$ Group of Applied Physics, University of Geneva, 1211 Geneva, Switzerland}
\ead{jiri.minar@unige.ch}
\begin{abstract}
We demonstrate fast and accurate control of the evolution of
collective atomic coherences in an Erbium doped solid using
external electric fields. This is achieved by controlling the
inhomogeneous broadening of Erbium ions emitting at 1536 nm using
an electric field gradient and the linear Stark effect. The
manipulation of atomic coherence is characterized with the
collective spontaneous emission (optical free induction decay)
emitted by the sample after an optical excitation, which does not
require any previous preparation of the atoms. We show that
controlled dephasing and rephasing of the atoms by the electric
field result in collapses and revivals of the optical free
induction decay. Our results show that the use of external
electric fields does not introduce any substantial additional decoherence and
enables the manipulation of collective atomic coherence with a
very high degree of precision on the time scale of tens of ns.
This provides an interesting resource for photonic quantum state
storage and quantum state manipulation.
\end{abstract}
\maketitle
\section{Introduction}
The coherent control of quantum systems plays a central role in
quantum information science and in quantum technology in general.
In particular, the coherent manipulation of collective atomic
coherences in material systems is crucial for applications in
photonic quantum storage \cite{Hammerer2008,Tittel2008} and in
ensemble based quantum computing \cite{Tordrup2008,Wesenberg2009}.
A promising route towards these applications is to use solid state
atomic ensembles implemented with rare-earth ion doped solids.

In a solid state environment, the optical atomic transitions of
the rare-earth impurities are inhomogeneously broadened
\cite{Macfarlane2002}. When atoms are excited, for example
following the absorption of a light pulse, the atomic dipoles
oscillate at different frequencies, leading to inhomogeneous
dephasing. In order to enable a constructive interference between
all the emitters that will lead to a collective re-emission of the
stored light, this dephasing must be controlled. The rephasing of
the dipoles can be triggered by optical pulses, as in traditional
photon echo techniques. These techniques, while very successful to
store classical light \cite{Mossberg1982,Lin1995} and as a tool
for high resolution spectroscopy
\cite{Macfarlane2002,Macfarlane1987a}, suffer from strong
limitations for the storage of single photons \cite{Ruggiero2009}.
Another possibility is to exploit the fact that some materials
exhibit permanent dipole moments, which give rise to a linear
Stark effect \cite{Macfarlane2007}. The frequency of the atoms can
then be controlled with a moderate external electric field. This
effect has been used in Stark-switched optical free induction decay \cite{Brewer1972,Shelby1978,Szabo1978} and in Stark modulated photon echoes \cite{Wang1992,Meixner1992,Graf1997,Graf1997a,Chaneliere2008} as a
tool for high resolution spectroscopy. The electric control of the
resonance frequency of the atoms is also the key resource of a
recently proposed modified photon echo protocol based on
controlled and reversible inhomogeneous broadening (CRIB)
\cite{Tittel2008,Moiseev2001,Kraus2006,Nilsson2005,Sangouard2007}.

The great advantage of using electric fields to manipulate atomic
coherences is that it does not change the population distribution
in the ground and excited states, contrary to optical rephasing
pulses. In combination with an optical transfer of population to a
long lived ground state, this enables in principle the long term
storage and retrieval of single photon fields with unit efficiency
and fidelity \cite{Kraus2006,Sangouard2007}. Quantum storage with
unit efficiency can also be achieved without the transfer to the
long lived ground state, using only electric fields for the light
retrieval \cite{Hetet2008,Longdell2008}. The first proof of
principle experiment of the CRIB protocol was performed in a Eu
doped crystal \cite{Alexander2006,Alexander2007JL} followed by
another demonstration in a Pr doped crystal \cite{Hetet2008}. The
maximal efficiency of the storage and retrieval is directly
proportional to the quality of the manipulation of the atomic
coherences. It is thus extremely important to have a good
characterization of the rephasing of the dipoles.

A CRIB experiment requires sophisticated optical pumping
techniques in order to first isolate a narrow absorption peak
within a large transparency window \cite{Tittel2008,Nilsson2005}. Moroever, it is not straightforward to
characterize the quality of the rephasing directly from a CRIB
experiment, since the efficiency of the storage also depends on
other parameters, such as available optical depth or quality of
the optical pumping for the preparation of the memory. In this
paper, we use a much simpler method to characterize the electric
manipulation of the atomic coherence. We propose to infer the
dynamics of the atomic coherences by studying the collective
spontaneous emission of light from the atoms (a phenomenon known
as optical free induction decay (FID) \cite{Brewer1972}). In
particular, we show that controlled dephasings and rephasings of
the atoms via the electric field result in collapses and revivals
of the FID. The observation of FID does not require any optical
preparation of the sample and  can thus be done with a relatively
simple experimental setup. In practice it is sufficient to excite
the atoms with a single optical pulse in resonance with the atomic
transition and to measure the collective emission of light after
the pulse. The FID is however strongly non linear with respect to
the excitation pulse intensity and vanishes in the limit of weak
excitation pulses \cite{Brewer1972,Afzelius2007}.

In this paper, we use Erbium ions doped into a Y$_{2}$SiO$_{5}$
crystal. This is an interesting system since Erbium ions have an
optical transition at the telecommunication wavelength of 1536 nm.
It could thus in principle enable the realization of a light
matter quantum interface between photons that can be transmitted
with low loss in optical fibers and stationary atoms in a solid.
Such a quantum memory at telecommunication wavelength would be
useful in the context of quantum repeaters
\cite{Duan2001,Sangouard2007a,Sangouard2008}. The spectroscopic
properties of Er$^{3+}$:Y$_2$SiO$_5$ have been extensively
studied, including optical coherence
\cite{Sun2002,Bottger2006,Bottger2006a,Bottger2009}, spectral
diffusion \cite{Bottger2006,Crozatier2007}, hyperfine structure
\cite{Guillot-Noel2006}, Zeeman relaxation lifetimes
\cite{Hastings-Simon2006}, Zeeman g factors \cite{Sun2008}, and
erbium host interactions \cite{Guillot-Noel2007}. Slow light has
also been achieved in this material using coherent population
oscillation \cite{Baldit2005}. Optical pumping techniques have
also been developed \cite{Lauritzen2008}, and a proof of principle
experiment of CRIB with weak light pulses at the single photon
level has been recently demonstrated in an
Er$^{3+}$Y$_{2}$SiO$_{5}$ crystal \cite{Lauritzen2009}.

\section{Control of collective atomic coherences using the linear Stark
effect} We now explain in more detail how the dephasing and
rephasing of atomic dipoles can be controlled via the linear Stark
effect, and how this affects the collective  emission of light
from the sample. In the presence of an external DC electric field,
the energy levels of an atom with a permanent dipole moment are
shifted by an amount proportional to the electric field. This
phenomenon is known as the linear DC Stark effect
\cite{Macfarlane2007}. If the dipole moments are different for
different electronic levels this shift leads to a shift in the
associated optical transition frequency. The linear DC Stark
effect can be observed in some RE-doped solids, where the ions possess
a permanent dipole moment induced by local electric fields due to
the crystal environment.
 The detuning of the
atom transition $\Delta$ due to the linear Stark effect can be
described as \cite{Hastings-Simon2006}
\begin{equation}
\Delta =\frac{\Delta \mu_e\chi \cos \theta}{\hbar}E\label{eq
Stark}
 \end{equation}
where $\Delta \mu_e$ is the difference between the permanent
dipole moments for the two states of the optical transition, E is
the applied electric field amplitude, $\chi=(\epsilon +2)/3$ is
the Lorentz correction factor, $\epsilon$ is the dielectric
constant of the sample and $\theta$ is the angle between
$\overrightarrow{\Delta \mu_e}$ and $\overrightarrow{E}$. Since
the crystal used in this experiment (Y$_2$SiO$_5$) is centro-symmetric, there are two classes of Erbium ions with
dipole moments pointing in opposite directions \cite{Macfarlane2007}.
This leads to the splitting of the resonance frequency of the
atoms when a homogeneous electric field is applied (pseudo-Stark splitting). Each frequency
is shifted by $\Delta$ or $-\Delta$ with respect to the
unperturbed absorption frequency $\omega^{at}$. If the electric
field intensity varies with the position in the sample, each atom
experiences a different Stark shift, which leads to an additional
inhomogeneous broadening.

When a light pulse is absorbed in an inhomogeneously broadened
sample, the atoms in resonance with the light will be excited.
While the excited atoms are in phase after the pulse is turned
off, they are in a superradiant state, and a strong collective
emission takes place in the forward mode defined by the input
pulse. This emission will then decay when the atoms dephase due to
inhomogeneous dephasing. The decay rate of the FID depends on the
spectral distribution of the excited atoms. It is thus possible to
accelerate the decay in a controlled way by broadening the
spectral distribution of the atoms with the linear Stark effect,
using an electric field gradient. The atoms are then no more in a
superradiant state and the collective emission is inhibited.
However, if no random phase is acquired during the process, it is
possible to undo the inhomogeneous dephasing due to the electric
field and to obtain a revival of the emission
\cite{Moiseev2001,Kraus2006,Nilsson2005}. This can be realized by
reversing the polarity of the electric field, which will reverse
the detuning of each atom. The phase evolution will now be
reversed and after a given time, all the atoms will be in phase
again, leading to a collective emission of light.

More formally, this process can be written as follows: suppose
that the phase of each atom evolves in time with frequency
$\omega^{at}$, i.e. we can write for the phase evolution of the
j-th atom: $e^{-i\omega^{at}_j t}$. An external electric field
will shift the frequency of the j-th atom by $\Delta_j$. Suppose
that electric field is turned on at time $t$=0 and turned off at
time $t=\tau$. The phase acquired by the atom is then
$e^{-i(\omega^{at}_j + \Delta_j) \tau}$. Now, suppose that instead
of turning off the electric field, it is switched to the opposite
polarity at time $t=\tau$. The frequency shift $\Delta_j$ then
becomes $-\Delta_j$ and the phase of the atom at time t is given
by \begin{equation}
 e^{-i(\omega^{at}_j + \Delta_j) \tau}
e^{-i(\omega^{at}_j - \Delta_j)(t-\tau)}. \label{eq Phase}
\end{equation}
 At a time $t=2\tau$ the externally introduced phase shifts $\Delta_j$
cancel out and the atomic phase oscillates again at the atomic
frequency $\omega^{at}_j$. If the optical transition has a natural
inhomogeneous broadening, the atomic evolution due to the
controlled dephasing and rephasing is superposed with the natural
evolution due to the inhomogeneous dephasing. Hence, in the case
of FID, the intensity of the light after the rephasing should
reach the intensity of the unperturbed FID signal.
\begin{center}
\begin{figure}[h!]
\includegraphics[width = 8 cm] {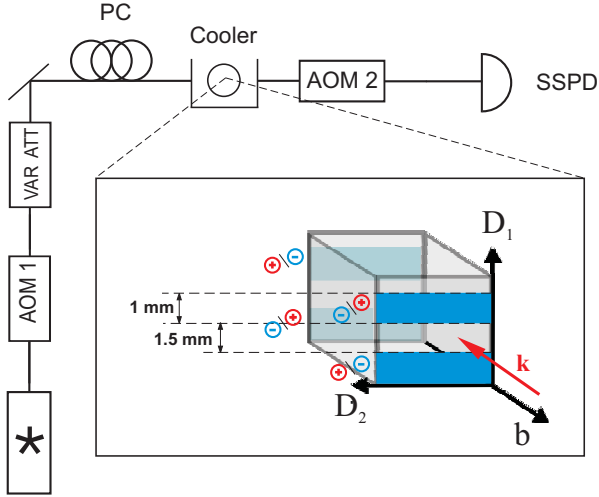}
\caption{(Color online) Experimental scheme used to demonstrate
the electric control of collective atomic coherences. Light pulses
created with an acousto-optic modulator (AOM1) are attenuated with
a fiber variable attenuator (VAR ATT) and focused through an
Er$^{3+}$:Y$_{2}$SiO$_{5}$ crystal cooled at 2.6 K in a pulse tube
cooler. The excitation pulses are then blocked by an optical gate
implemented with a fiber acousto-optic modulator (AOM2) and the
weak FID signal at the single photon level is detected with a
superconducting single photon detector (SSPD). The inset shows the
crystal with the quadrupole configuration of electrodes that
produce the electric field gradient along the light propagation
direction. } \label{FID_schema_SSPD}
\end{figure}
\end{center}
\section{The experiment}

In the experiment we used a Y$_2$SiO$_5$ crystal doped with erbium
ions Er$^{3+}$ (with 10 ppm concentration). The atoms were excited
on the transition $^{4}I_{15/2} - ^{4}I_{13/2}$ at the telecom
wavelength of 1536 nm \cite{Bottger2006}. The Y$_{2}$SiO$_{5}$
crystal has three mutually perpendicular optical-extinction axes
labelled D$_{1}$, D$_{2}$, and b. The direction of light
propagation $\vec{k}$ is along the b axis.  The dimensions of the
Er$^{3+}$:Y$_{2}$SiO$_{5}$ crystal are 6mm x 3.5mm x 4 mm along
the b, D$_{1}$, D$_{2}$ axis, respectively. The crystal was cooled
to 2.6 K in a pulse tube cooler (Oxford Instruments). The optical
absorption depth of the crystal is $\alpha L=\,$2. In order to create the
electric field gradient, we implemented a quadrupole scheme
\cite{Alexander2006} using four electrodes attached directly to
the crystal, perpendicular to the D$_1$ axis (see inset of
Fig. \ref{FID_schema_SSPD}). Such a configuration creates an
electric field in the D$_1$ direction which changes linearly along
the axis parallel to the light propagation. The electrodes were
thin aluminium stripes, each of 1 mm width and spaced by 1.5
mm. To switch the electric field we used a fast electrical switch
with a switching time of 10 ns and minimal/maximal voltage
-100/100 V.


The experimental setup is shown in Fig. \ref{FID_schema_SSPD}. The
light source was a free running external cavity diode laser
(Toptica) at 1536 nm. The light was amplitude modulated by an
acousto-optic modulator in order to create the excitation pulses
at a repetition rate of 10 kHz, with duration 3 $\mu s$. The light
was then coupled to a single mode optical fiber and passed through
a variable fiber attenuator, before being focused in free space
through the crystal in the cryostat. After the crystal, the light
was again coupled into a single mode fiber and sent through a
fiber coupled AOM that served as optical gate in front of the
detector to block the excitation pulses. The measurements were
made in the low excitation regime (with an excitation pulse area
$\ll \pi$, typically with about 10$^6$ photons in the excitation
pulses). As the amplitude of the FID signal is strongly non linear
with respect to the excitation intensity, the FID signal was
extremely weak (about 50 photons) and was detected with a
superconducting single photon detector (SSPD)
\cite{Gol'tsman2001}.

\section{Experimental collapses and revivals of collective
spontaneous emission}
\begin{figure}[h!]
\begin{center}
\epsfxsize=1\textwidth \epsfbox{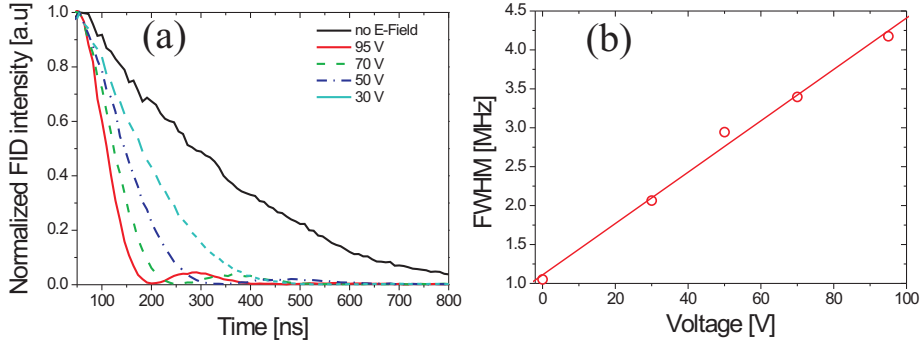}
\end{center}\caption{(Color online) (a) Decay of the FID for different values of
electric fields. The electric field is switched on after the pulse
and kept constant afterwards. (b) Broadening of the excited atoms
as a function of the voltage applied on the electrodes. The value
plotted is the full width at half maximum (FWHM) of the spectral
distribution given by the Fourier transform of the decay curves in
(a).} \label{fig decay}
\end{figure}

\begin{center}
\begin{figure}[h!]
\includegraphics[width = 12 cm] {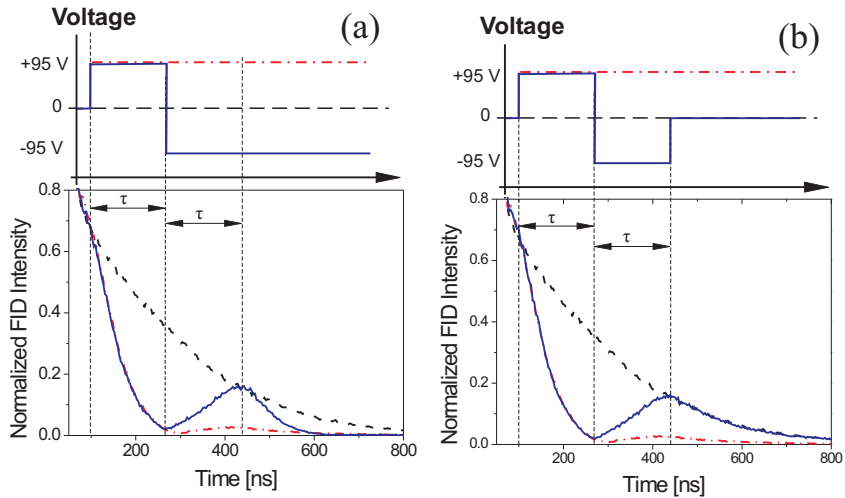}
\caption{(Color online) Collapse and revival of collective
emission for different electric field sequences. The unperturbed
FID signal is represented by the dashed line. The dashed-dotted
curve represents the damped FID signal when the electric field is
not switched. The voltage applied on the electrodes is $\pm$ 95 V.
(a) Temporary revival obtained when the polarity of the electric
field is reversed at time $\tau$ and the field remains constant
afterwards. (b) Revival obtained when the polarity of the electric
field is reversed at time $\tau$ and the field is turned off at
time 2$\tau$. The revived signal then follows the unperturbed FID
signal.} \label{FID_manip_SSPD}
\end{figure}
\end{center}
\begin{center}
\begin{figure}[h!]
\includegraphics[width = 8 cm] {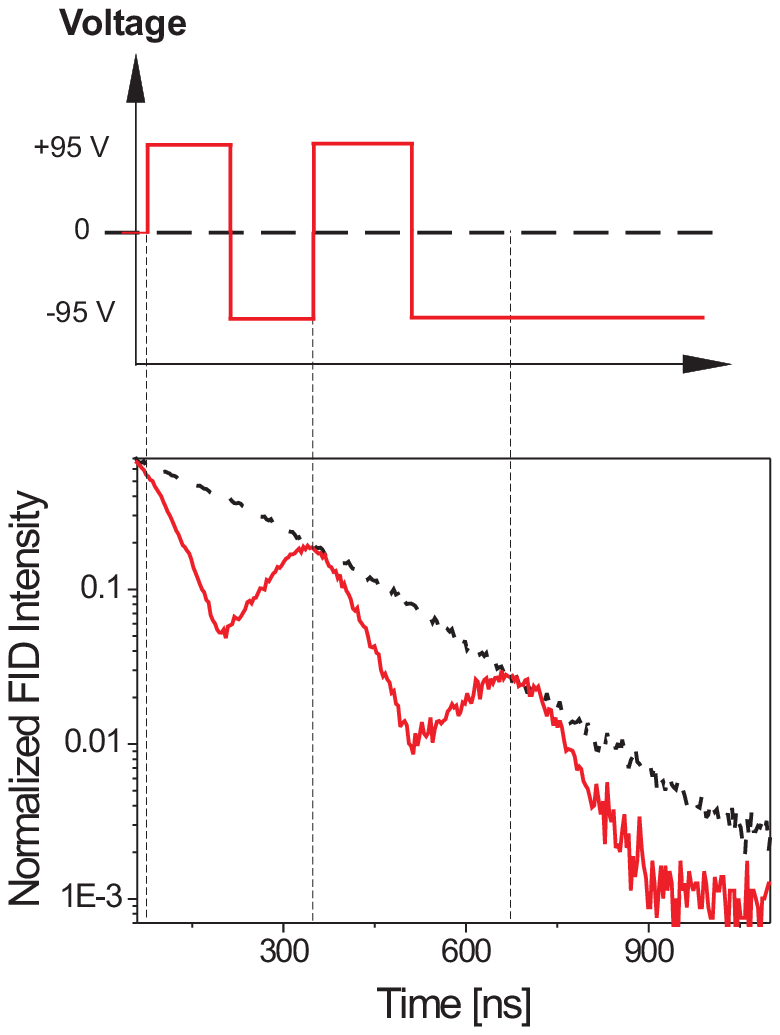}
\caption{(Color online) Multiple collapse and revival obtained when the polarity
of the electric field is reversed repeatedly. The voltage applied
on the electrodes is $\pm$ 95 V.} \label{Double_rev_SSPD}
\end{figure}
\end{center}
We now present experimental results of the observation of collapse
and revival of collective emission. In a first experiment, we
apply a positive Stark pulse after the end of the optical pulse
and we observe the decay rate of the FID for different values of
electric fields. The results are shown in Fig. \ref{fig decay}a.
We see that the decay becomes faster when the electric field
increases, due to the applied broadening. For high value of
electric fields, we also observe a small revival after the FID
goes to zero. This is due to the spectral distribution of the
induced broadening, which approaches a square shape when the field
increases. In Fig. \ref{fig decay}b, we plot the full width at
half maximum of the spectral distribution, given by the Fourier
transform of the FID decay. The linear dependance confirms that
the spectral distribution of the excited atoms is proportional to
the applied electric field.
\begin{figure}[h]
\begin{center}
\includegraphics[width = 10 cm] {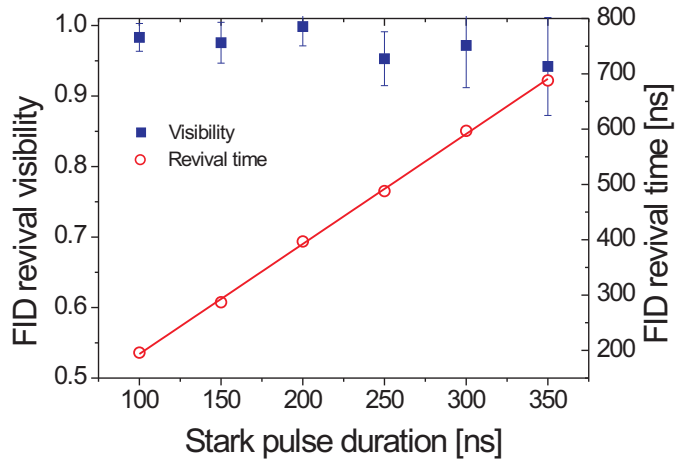}
\end{center}
\caption{(Color online) Visibility of the revival as a function of
the first Stark pulse duration $\tau$. In order to obtain an
estimation of the visibility free of intensity fluctuations, the
measurement sequence is composed of two optical pulses. The
electric field sequence is applied to one pulse, while the other
pulse serves as reference (unperturbed FID). For long Stark pulses
durations, the visibility is more difficult to be estimated due to
the low count rate. The error bars correspond to the statistical
uncertainty of photons counts. The time of revival as a function
of $\tau$ with a slope of 1.99$\pm$ 0.02 is also plotted.}
\label{fig revival}
\end{figure}
We then show that the controlled dephasing is reversible, by
switching the polarity of the Stark pulse after a time $\tau$ and
observing the revival of collective emission due to the rephasing
of the atoms. Different sequences of electrical pulses can be used
in order to manipulate the atomic coherence in the desired manner.
In \fig{\ref{FID_manip_SSPD} we present two examples of such
sequences. If the electric field is kept constant after the
switching, the atoms are in phase again after a time $2\tau$ and
we observe a temporary revival of the collective emission at this
time (Fig. \ref{FID_manip_SSPD}a). However, if the field is
switched off to zero at the time $2\tau$, the controlled phase
evolution is frozen and the natural inhomogeneous dephasing
governs the evolution. In that case, the revived signal follows
the unperturbed FID signal (Fig. \ref{FID_manip_SSPD}b). The
dashed curve represents the original unperturbed FID signal. If
the electric field is switched repeatedly, it is also possible to
induce multiple revivals, as shown in Fig. \ref{Double_rev_SSPD}.

In all cases it is clearly observed that the quality of rephasing
is excellent, i.e. the revived FID signal reaches almost the
unperturbed signal. This suggests that the process of manipulating
atomic coherence using electric fields does not introduce any
substantial noise which would cause additional decoherence. In
order to have a more quantitative estimation of the quality of the
rephasing, we induced temporary revivals at different times. We
varied the duration of the first Stark pulse, and hence the moment
of the switching $\tau$. We measured the visibility $V$ of the
revival as a function of $\tau$. $V$ is defined as
$V=I_{revival}/I_{FID}$, where $I_{revival}$ is the maximal
intensity of the revival and $I_{FID}$ is the intensity of the
unperturbed FID at the corresponding time. In order to have an
accurate estimation of the visibility, it is important that the
revival and the reference measurement (unperturbed) FID are taken
in the same experimental conditions. Since the FID intensity is
extremely dependent on the laser frequency and intensity, and that
the typical measurement times are a few hundred seconds, it is
difficult to ensure the same experimental conditions for the two
measurements. In order to overcome this problem, we implemented a
measurement sequence with two subsequent optical excitation pulses
within 10$\mu s$. The electric field was applied for one pulse,
while the other pulse served as reference. The result is shown in
Fig. \ref{fig revival}. We observe that the visibility stays
constant above 0.95 within the error bars for all delays. In Fig.
\ref{fig revival}, we also plot the time of the revivals as a
function of $\tau$. We measure a slope of $1.99\pm0.02$, which
confirms that the revivals happen after a time $2\tau$.

These results show that the dephasing and rephasing of the
collective atomic coherence can be controlled to a very high
degree. This is a crucial capability for applications in photonic
quantum storage based on controlled reversible inhomogeneous
broadening. More generally, this ability to switch on and off at
will the collective emission of light from the sample is an
interesting resource for quantum state engineering and quantum
state manipulation.\\


\section{Conclusion}

In conclusion, we have shown that the collective atomic coherence
of an ensemble of erbium ions embedded in a solid state matrix can
be controlled to a high degree on the time scale of tens of
nanoseconds using external electric fields. Controlled
inhomogeneous dephasing and rephasing was implemented using a
reversible linear gradient of electric field on the crystal.

We used  optical free induction decay to test our capacity to
manipulate the atomic coherence. In particular, we showed that the
controlled dephasing and rephasing of the atomic dipoles results
in collapses and revivals of the collective emission of light from
the sample. The experimental results show that the use of the
electric field does not introduce any substantial additional decoherence and
enables us to manipulate the atomic coherence efficiently on the
time scale of tens of ns. It thus provides a useful resource in
quantum information science, in particular for quantum storage
applications.

We would like to thank C. Barreiro  and J-D. Gautier for technical support.
Financial support from the swiss NCCR Quantum Photonics, from the EU integrated project QAP and from the ERC-AG Qore is acknowledged.\\
~\\

\bibliography{mybibCS}

\begin{thebibliography}{10}

\bibitem{Hammerer2008}
K. Hammerer, A. S{\o}rensen, and E. Polzik, arXiv:0807.3358  (2008).

\bibitem{Tittel2008}
W. Tittel {\it et~al.}, arXiv:0810.0172  (2008).

\bibitem{Tordrup2008}
K. Tordrup, A. Negretti, and K. Molmer, Phys. Rev. Lett. {\bf 101},  040501
  (2008).

\bibitem{Wesenberg2009}
J. Wesenberg {\it et~al.}, arXiv:0903.3506v1  (2009).

\bibitem{Macfarlane2002}
R.~M. Macfarlane, J. Lumin. {\bf 100},  1  (2002).

\bibitem{Mossberg1982}
T.~W. Mossberg, Opt. Lett. {\bf 7},  77  (1982).

\bibitem{Lin1995}
H. Lin, T. Wang, and T.~W. Mossberg, Opt. Lett. {\bf 20},  1658  (1995).

\bibitem{Macfarlane1987a}
R. Macfarlane and R. Shelby,  in {\em Coherent Transients And Holeburning
  Spectroscopy In Rare Earth Ions In Solids; Spectroscopy Of Crystals
  Containing Rare Earth Ions}, edited by A. Kaplyankii and R. Macfarlane
  (Elsevier Science Publishers, Amsterdam, Netherlands, 1987).

\bibitem{Ruggiero2009}
J. Ruggiero, J.-L. Le~Gou{\"{e}}t, C. Simon, and T. Chaneli{\`e}re, Phys. Rev.
  A {\bf 79},  053851  (2009).

\bibitem{Macfarlane2007}
R.~M. Macfarlane, Journal of Luminescence {\bf 125},  156  (2007).

\bibitem{Brewer1972}
R.~G. Brewer and R.~L. Shoemaker, Phys. Rev. A {\bf 6},  2001  (1972).

\bibitem{Shelby1978}
R.~M. Shelby and R.~M. Macfarlane, Optics Communications {\bf 27},  399
  (1978).

\bibitem{Szabo1978}
A. Szabo and M. Kroll, Optics Letters {\bf 2},  10  (1978).

\bibitem{Wang1992}
Y.~P. Wang and R.~S. Meltzer, Phys. Rev. B {\bf 45},  10119  (1992).

\bibitem{Meixner1992}
A.~J. Meixner, C.~M. Jefferson, and R.~M. Macfarlane, Phys. Rev. B {\bf 46},
  5912  (1992).

\bibitem{Graf1997}
F.~R. Graf, A. Renn, U.~P. Wild, and M. Mitsunaga, Phys. Rev. B {\bf 55},
  11225  (1997).

\bibitem{Graf1997a}
F.~R. Graf {\it et~al.}, Opt. Lett. {\bf 22},  181  (1997).

\bibitem{Chaneliere2008}
T. Chaneli{\`e}re {\it et~al.}, Phys. Rev. B {\bf 77},  245127  (2008).

\bibitem{Moiseev2001}
S.~A. Moiseev and S. Kr\"oll, Phys. Rev. Lett. {\bf 87},  173601  (2001).

\bibitem{Kraus2006}
B. Kraus {\it et~al.}, Phys. Rev. A {\bf 73},  020302  (2006).

\bibitem{Nilsson2005}
M. Nilsson and S. Kr\"oll, Optics Communications {\bf 247},  393  (2005).

\bibitem{Sangouard2007}
N. Sangouard, C. Simon, M. Afzelius, and N. Gisin, Phys. Rev. A {\bf 75},
  032327  (2007).

\bibitem{Hetet2008}
G. H\'etet {\it et~al.}, Phys. Rev. Lett. {\bf 100},  023601  (2008).

\bibitem{Longdell2008}
J.~J. Longdell, G. H{\'e}tet, P.~K. Lam, and M.~J. Sellars, Phys. Rev. A {\bf
  78},  032337  (2008).

\bibitem{Alexander2006}
A.~L. Alexander, J.~J. Longdell, M.~J. Sellars, and N.~B. Manson, Phys. Rev.
  Lett. {\bf 96},  043602  (2006).

\bibitem{Alexander2007JL}
A.~L. Alexander, J.~J. Longdell, M.~J. Sellars, and N.~B. Manson, Journal of
  Luminiscence {\bf 127},  94  (2007).

\bibitem{Afzelius2007}
M. Afzelius {\it et~al.}, New Journal of Physics {\bf 9},  413  (2007).

\bibitem{Duan2001}
L.-M. Duan, M.~D. Lukin, J.~I. Cirac, and P. Zoller, Nature {\bf 414},  413
  (2001).

\bibitem{Sangouard2007a}
N. Sangouard {\it et~al.}, Phys. Rev. A {\bf 76},  050301 (R)  (2007).

\bibitem{Sangouard2008}
N. Sangouard {\it et~al.}, Phys.Rev.A {\bf 77},  062301  (2008).

\bibitem{Sun2002}
Y. Sun {\it et~al.}, J. Lumin. {\bf 98},  281  (2002).

\bibitem{Bottger2006}
T. B\"{o}ttger, Y. Sun, C.~W. Thiel, and R.~L. Cone, Phys. Rev. B {\bf 74},
  075107  (2006).

\bibitem{Bottger2006a}
T. B\"{o}ttger, C.~W. Thiel, Y. Sun, and R.~L. Cone, Phys. Rev. B {\bf 73},
  075101  (2006).

\bibitem{Bottger2009}
T. B{\"{o}}ttger, C.~W. Thiel, R.~L. Cone, and Y. Sun, Phys. Rev. B {\bf 79},
  115104  (2009).

\bibitem{Crozatier2007}
V. Crozatier {\it et~al.}, Journal of Luminescence {\bf 127},  65  (2007).

\bibitem{Guillot-Noel2006}
O. Guillot-No\"{e}l {\it et~al.}, Phys. Rev. B {\bf 74},  214409  (2006).

\bibitem{Hastings-Simon2006}
S.~R. Hastings-Simon {\it et~al.}, Optics Communications {\bf 266},  716
  (2006).

\bibitem{Sun2008}
Y. Sun, T. B\"ottger, C.~W. Thiel, and R.~L. Cone, Phys. Rev. B {\bf 77},
  085124  (2008).

\bibitem{Guillot-Noel2007}
O. Guillot-No{\"{e}}l {\it et~al.}, Phys. Rev. B {\bf 75},  205110  (2007).

\bibitem{Baldit2005}
E. Baldit {\it et~al.}, Phys. Rev. Lett. {\bf 95},  143601  (2005).

\bibitem{Lauritzen2008}
B. Lauritzen {\it et~al.}, Phys. Rev. A {\bf 78},  043402  (2008).

\bibitem{Lauritzen2009}
B. Lauritzen {\it et~al.}, in preparation  (2009).

\bibitem{Gol'tsman2001}
G.~N. Gol'tsman {\it et~al.}, Appl. Phys. Lett. {\bf 79},  705  (2001).

\end{thebibliography}
\bibliographystyle{prsty}

\end{document}